\documentclass[english,aps,prl,twocolumn,showpacs,superscriptaddress,groupedaddress,footinbib,reprint,noshowpacs]{revtex4}
\usepackage{graphicx}
\usepackage{amsmath}
\usepackage{comment}
\usepackage{amssymb}
\newcommand{\angstrom}{\textup{\AA}}
\usepackage{subfigure}

\usepackage{subfigure}
\usepackage{color}
\usepackage{soul}

\begin{document}

\title{Microscopic mechanism of thermomolecular orientation and polarization} 

\author{Alpha A. Lee}
\email{alphalee@g.harvard.edu}
\affiliation{School of Engineering and Applied Sciences, Harvard University, Cambridge, MA 02138}

\begin{abstract}

Recent molecular dynamics simulations show that thermal gradients can induce electric fields in water that are comparable in magnitude to electric fields seen in ionic thin films and biomembranes. This surprising non-equilibrium phenomenon of thermomolecular orientation is also observed more generally in simulations of polar and non-polar size-asymmetric dumbbell fluids. However,  a microscopic theory linking thermomolecular orientation and polarization to molecular properties is yet unknown.  Here, we formulate an analytically solvable microscopic model of size-asymmetric dumbbell molecules in a temperature gradient using a mean-field, local equilibrium approach. Our theory reveals the relationship between the extent of thermomolecular orientation and polarization, and molecular volume, size anisotropy and dipole moment. Predictions of the theory agree quantitatively with molecular dynamics simulations. Crucially, our framework shows how thermomolecular orientation can be controlled and maximized by tuning microscopic molecular properties.

\end{abstract} 

\makeatother
\maketitle

Non-equilibrium effects play an important role in processes relevant to biology, chemistry, physics and materials science. In particular, temperature gradients often trigger a plethora of coupling effects. For example, thermal gradients trigger mass transport, commonly known as the Soret effect \cite{mazur1963non,rahman2014thermodiffusion}. Colloids in a suspension tend to move towards the colder or hotter region, and the direction and magnitude of the Soret response depend on molecular properties such as charge and size \cite{piazza2008thermophoresisjpcm,wurger2010thermal}. Similarly, mixtures of molecules separate in response to a thermal gradient, with molecules moving to the hotter or colder region depending on mass, moment of inertia and intermolecular interactions \cite{haase1950thermodynamischphanomenologischen,kempers1989thermodynamic,shukla1998new,debuschewitz2001molecular,artola2007microscopic}. As such, the Soret effect allows efficient separation of mixtures of nanoscopic particles and biomolecules \cite{piazza2008thermophoresis,duhr2006molecules,richter2008magnesium,wienken2010protein,huang2010isotope,dominguez2011soret,maeda2012effects,lervik2014sorting}. 

\begin{figure}
\centering
\includegraphics[scale=0.5]{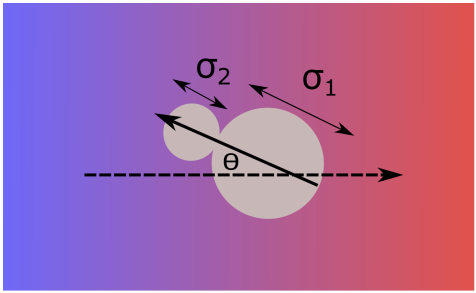}
\caption{Schematic sketch of the system under consideration: a model anisotropic molecule consisting of two touching spheres of diameters $\sigma_1$ and $\sigma_2$ in a thermal gradient. The dashed line indicates direction of the heat flux.}
\label{schematic}
\end{figure}

Recent molecular dynamics simulations have shown that, in addition to thermophoretic mass transport, thermal gradients can also trigger a preferential orientation if the molecule is anisotropic (the ``thermomolecular orientation'' effect) \cite{romer2012thermomolecular}. Uncharged diatomic molecules with a larger ``head'' and a smaller ``tail'' orient  in response to a thermal gradient, with the ``fatter'' end of the molecule pointing towards the hot region (see Figure \ref{schematic}). If the molecule is polar, this preferential orientation can induce a significant electric field. For liquid water, the thermal polarization coefficient can be  $\sim 10^{2}- 10^{4} \mathrm{V/m}$ for temperature gradients of the order of $\sim 10^6- 10^8 \mathrm{K/m}$; this coefficient is a strong function of the thermodynamic state and is largest near criticality \cite{bresme2008water,iriarte2016rich}. The magnitude of this thermal polarization coefficient is comparable to electric fields in ionic thin films and biomembranes, and the large thermal gradient can be generated \emph{in vivo} by electromagnetic heating of nanoparticles introduced into biological tissues \cite{jain2007nanoparticles}. Therefore, this effect of hot water polarization may be relevant to proposals of destroying cancer cells with nanoparticles and radiation sources \cite{bresme2008water}. Moreover, a recent theory \cite{frenkel2016hot} suggests that localized heat sources may interact as magnetic monopoles (charges) if the heat-polarization coupling is realised in ferrofluids (dipolar fluids). 

However, a microscopic theory that relates molecular properties, such as the degree of anisotropy and dipole moment, to the extent of thermomolecular orientation and the magnitude of the electric field generated is yet unknown. Prior studies of thermomolecular orientation and polarization rely on macroscopic non-equilibrium thermodynamic theory, whereby linear flux-force relations is posited with the Onsager coefficients satisfying symmetry laws \cite{bresme2008water,romer2012thermomolecular}. Those pioneering works have revealed the dependence of the response field on the local temperature and thermal gradient: In a non-polar liquid, $\left<\cos \theta_x \right>$ is directly proportional to the thermal gradient and inversely proportional to the local temperature, where $\theta_x$ is the angle between the $x$-axis and the orientation vector $\mathbf{n}$ of a molecule (defined to be a unit vector in the direction of the molecular axis pointing from the larger end to the smaller end), and the thermal gradient is applied in the $x$-direction \cite{romer2012thermomolecular}. For polar liquids, the induced electric field $\mathbf{E}$ scales inversely with temperature and linearly with thermal gradient \cite{bresme2008water}.  However, the constants of proportionality in those relationships, which were left as phenomenological coefficients, depend on molecular properties and, crucially, determine the magnitude of the response field. 


In this Letter, we will explore the physical mechanism of this novel non-equilibrium phenomena by developing a microscopic theory for thermomolecular orientation and polarization. We will first derive the relationship between microscopic molecular properties and the thermomolecular orientation for non-polar dumbbell molecules using a local equilibrium approach. We will then extend the theory to consider the thermomolecular polarization of polar dumbbell molecules, and reveal a surprising non-monotonic dependence of the polarization on the dipole moment. Our results will be verified through comparison with non-equilibrium molecular dynamics simulations. Underlying our approach, we argue that the microscopic mechanism of thermomolecular orientation and polarization is an extension of the Soret effect: the torque exerted on an anisotropic molecule is due to a stronger Soret force pushing the larger end of the molecule toward the warmer region than the Soret force acting on the smaller end. 

Consider first a simpler problem of thermophoresis of hard spheres suspended in a fluid. We will use the local equilibrium approach, and assume that the equilibrium free energy and chemical potential continue to be valid outside equilibrium, but with the temperature replaced by the local temperature \cite{eastman1926thermodynamics, eastman1928theory,prigogine1950recherches,mortimer1980elementary,artola2008new,duhr2006molecules}. The local equilibrium picture in thermophoresis holds as long as molecules are in mechanical equilibrium, which is almost always the case due to large viscous dissipation \cite{astumian2006unreasonable, astumian2007coupled}. A simple model is that the interfacial (solvation) energy is proportional to the exposed surface area ,
\begin{equation}
G_{\mathrm{solv}} = - k_B T s \sigma^2, 
\label{g_solv}
\end{equation}
where $\sigma$ is the hard sphere diameter, and $s$ is a positive constant if we assume that the solute is solvophilic (hence demixing is thermodynamically unfavourable) and negative vice versa. As thermophoresis is an interfacial phenomenon, the bulk free energy ($\sim \sigma^3$) is irrelevant. A thermal gradient therefore creates a spatially varying interfacial free energy, and the molecules moves along this chemical potential gradient to minimise its free energy. Substituting Equation (\ref{g_solv}) into the well-known relation in local equilibrium theory between Soret coefficient and free energy, $S_T = (\mathrm{d}G/\mathrm{d}T)/(k_B T)$ \cite{eastman1926thermodynamics, eastman1928theory}, yields 
\begin{equation}
S_T \propto - s \sigma^2.
\end{equation}
The linear dependence of the Soret coefficient on molecular area agrees with experimental results \cite{duhr2006thermophoretic}. We note that there are experimental studies with colloids suggesting that the Soret coefficient scales linearly with colloid \emph{diameter} \cite{braibanti2008does,piazzaR2008thermophoresis}. This discrepancy has been analysed in \cite{wurger2013soret}:  $S_T \propto \sigma$ in the regime where thermophoresis is driven by boundary layer flow near the colloid surface, which is the case if $\sigma \gg \lambda$, with $\lambda$ the range of interaction between the molecule/colloid and the solvent. In the opposite small-molecule regime of $\sigma \ll \lambda$, the physics is dominated by the solvation enthalpy, which scales with the molecular/colloidal area. As such, the scaling of the Soret coefficient depend on the solvent-solute interactions. In the case of a small molecule in a solvent that comprise other small molecules, which we will consider shortly, one expects the range of the intermolecular interactions to be larger than the molecular size, i.e. $\sigma \ll \lambda$, thus the Soret effect is enthalpy-dominated. Focusing on molecular mixtures rather than colloidal systems, non-equilibrium molecular dynamics simulations of binary Lennard-Jones systems suggest that the Soret coefficient can be fitted to a polynomial with terms linear and quadratic in particle diameter \cite{reith2000nature,bordat2001influence}. Nonetheless the size ratios investigated in those pioneering works were insufficient to conclusively determine the scaling of the Soret coefficient with particle diameter for molecular fluids. Future works which investigate a larger range of size ratios, as well as different range and strength of interparticle interactions, are required to verify the linear and quadratic scaling regime of the Soret coefficient. 


A single hard sphere evidently cannot display thermo-molecular orientation. The crucial ingredient missing is shape anisotropy, which can be realized by joining two touching hard spheres of different diameters  (see Figure \ref{schematic}). We note that simulations of thermomolecular orientation considered a pure fluid of size-asymmetric dumbbells \cite{bresme2008water,romer2012thermomolecular,daub2014thermo}, rather than dumbbells suspended in a background fluid. In this respect, thermomolecular orientation is different to thermophoresis of colloids. Nonetheless, treating the molecules as a continuum, the quantity in a pure system analogous to the interfacial energy (\ref{g_solv}) is the solvation energy of the molecule in a bath of the same molecule. We will use this analogy in the remainder of this paper. We emphasise that this analogy is an approximation and an extrapolation as the separation of length scale between colloids/macromolecules underpinning the solvation energy picture \cite{duhr2006molecules} does not apply in a molecular fluid. However, comparison of our theoretical predictions with results from molecular dynamics simulations will show that our approximation is indeed reasonable. Moreover, the connection between the Soret effect and thermomolecular orientation was shown in molecular dynamics simulations \cite{romer2012thermomolecular}, where the thermomolecular orientation of the dumbbell molecule was found to correlate with the Soret coefficient of the corresponding binary mixture. 

To apply the local equilibrium approach to thermomolecular orientation, we assume that the total free energy of the dumbbell molecule is simply the sum of $G_{\mathrm{solv}}$ for the large and small spheres, but crucially with different temperatures because the spheres are in a thermal gradient
\begin{equation}
G_{\mathrm{tot}} = -k_B s \left( T_1 \sigma_1^2 + T_2 \sigma_2^2 \right),  
\end{equation} 
where $T_{1,2}$ is the local temperature at sphere $1$/$2$. Physically, the lengthscale of temperature variation is much larger than the length of the dumbbell. Therefore, $T_{1,2}$ is related to the local temperature $T$ and gradient $\nabla_{x} T$ at the midpoint of the dumbbell via 
\begin{align}
T_{1} &\approx T + \frac{l}{2} \cos \theta \nabla_{x} T, \nonumber \\ 
T_{2} &\approx T - \frac{l}{2} \cos \theta \nabla_{x} T, 
\label{lin_approx_temp}
\end{align}
where $l = \sigma_1 + \sigma_2$ is the length of the dumbbell. (For linear temperature gradients, Equation (\ref{lin_approx_temp}) holds identically.) The average orientation $\left< \cos \theta_x\right> $ can be computed by a Boltzmann average assuming local equilibrium 
\begin{align}
\left< \cos \theta_x\right> &= \frac{ \int_0^{\pi} \cos \theta \sin \theta \; e^{- \frac{G_{\mathrm{tot}}}{k_B T}} \mathrm{d} \theta}{\int_0^{\pi}  \sin \theta \; e^{- \frac{G_{\mathrm{tot}}}{k_B T}} \mathrm{d} \theta}, \nonumber \\ 
&= \frac{ \int_0^{\pi} \cos \theta \sin \theta \; \exp\left[ -\frac{s l}{2 T}  \cos \theta (\sigma_2^2 - \sigma_1^2) \nabla_{x} T \right] \mathrm{d} \theta}{ \int_0^{\pi} \sin \theta \; \exp\left[ -\frac{s l}{2 T}  \cos \theta (\sigma_2^2 - \sigma_1^2) \nabla_{x} T \right] \mathrm{d} \theta}. 
\label{boltzmann_integral}
\end{align} 
In the linear response regime, $\nabla_x T \ll1 $, and the integrals in Equation (\ref{boltzmann_integral}) can be evaluated to give 
\begin{equation} 
\left< \cos \theta_x\right> \approx \frac{1}{6} s l \frac{\nabla_{x} T}{T} \sigma_2^2 \left[\left(\frac{\sigma_1}{\sigma_2}\right)^2-1 \right].  
\label{ave_orien}
\end{equation} 
Assuming a fixed molecule length $l$, Equation (\ref{ave_orien}) shows that $\left< \cos \theta_x\right>$ is related to the size asymmetry ratio $\chi = \sigma_2/\sigma_1 $ via 
\begin{equation}
\left< \cos \theta_x\right> = \frac{s l^3}{6} \frac{\chi-1}{\chi+1} \frac{\nabla_{x} T}{T}. 
\label{cent_eq}
\end{equation} 

\begin{figure}
\includegraphics[scale=0.3]{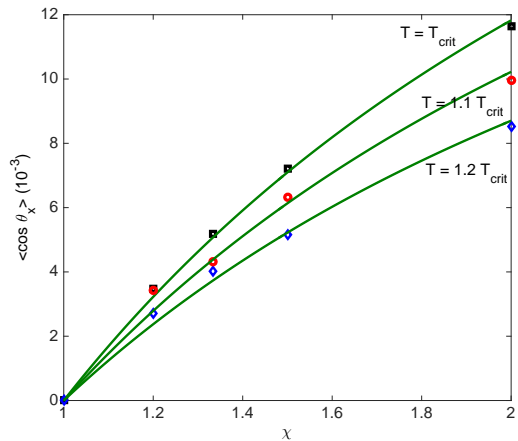}
\caption{Comparison of our theory with non-equilibrium molecular dynamics simulations of Lennard-Jones dumbbells \cite{romer2012heat}. In those simulations the gradient $\nabla_{x} T = 1 K/ \angstrom$ is kept constant, and the temperature is scaled to the corresponding critical value for given $\chi$.  The function $A (\chi-1)/(\chi+1)$ (green solid line, c.f. Equation (\ref{cent_eq})) can be fitted (using $A$) to simulation data at density $\rho = 2.5 \rho_{\mathrm{crit}}$ (open squares, circle and diamond) with $A = 0.035$ ($T=T_{\mathrm{crit}}$), $A=0.031$ ($T=1.1 \; T_{\mathrm{crit}}$) and $A=0.026$ ($T=1.2 \; T_{\mathrm{crit}}$). Note that the average orientation for $\chi=1$ is used as the zero baseline to calibrate uncertainties in simulations. }
\label{fit_sim} 
\end{figure}

Equation (\ref{cent_eq}) is a central result of this Letter. It shows how the magnitude of thermomolecular orientation depends on the size asymmetry of the molecule and molecular volume. The only unknown in Equation (\ref{cent_eq}) is the relationship between $s$, the solvation energy (in units of thermal energy) per exposed surface area, and $\chi$. This relationship can be estimated by noting that the phase diagram of size-asymmetric hard dumbbells collapse onto a single one when the temperature and density are scaled with the corresponding critical values \cite{romer2012heat}. As the critical temperature is a measure of the interactions that render the fluid state stable, we will assume that $s \propto T_{\mathrm{crit}}$. Figure \ref{fit_sim} shows that the average orientation obtained from simulations \cite{romer2012heat} quantitatively agrees with the scaling relationship $\left< \cos \theta_x\right> \propto (\chi-1)/(\chi+1) $ when the temperature is expressed in terms of the critical temperature. Increasing the temperature by a factor of $1.2$ increases the prefactor by a factor of $1.3$, close to the expected scaling. Moreover, the numerical value of $s$ agrees with the interpretation that it is the solvation energy per surface area (c.f. Equation (\ref{g_solv})): from the $\chi=2$ simulation with $T= T_{\mathrm{crit}}$, where the response is the most pronounced, the prefactor obtained from fitting (Figure \ref{fit_sim}) translates to $k_B T s \approx 0.05 k_B T / \angstrom^2$. Thus the solvation energy for a sphere with diameter $\sigma = 3 \angstrom$, the mean sphere diameter, is $G_{solv} \approx - k_B T s \sigma^2 \approx - 0.5 k_B T$. This agrees with depth of the Lennard-Jones energy minimum $\epsilon/k_B T \approx 0.7$ of the system (as the Lennard-Jones energy is short ranged, the solvation energy is dominated by interactions with the first solvation shell, which is the same order of magnitude as the interaction energy between two Lennard-Jones particle).  In addition, our model predicts that both the Soret coefficient $S_T$ and the average orientation $\left< \cos \theta_x\right>$ are linearly proportional to the solvation energy per molecular area $s$, agreeing with linear relationship between the Soret coefficient and $\left< \cos \theta_x\right>$ reported in molecular simulations\cite{romer2012heat}. 

Importantly, Equation (\ref{cent_eq}) predicts that the orientation effect depends rather weakly on $\chi$ for large $\chi$, yet scales linearly with $l^3$, the molecular volume. Those scalings suggest that in order to obtain the strongest orientation effect, it is more important to use large molecules than ones with large size asymmetry. 

For \emph{dipolar}, size-asymmetric dumbbells, thermomolecular orientation generates a concomitant electric field. This electric field can be estimated using a self-consistent mean-field approximation. Analogous to the approach above, we first note that the free energy of a size-asymmetric dipole of dipole moment $\mu$ in a thermal gradient with an electric field $E$ in the direction of the thermal gradient (which will be computed later by imposing self-consistency) is given by
\begin{equation}
G_{\mathrm{tot}} = -k_B s \left( T_1 \sigma_1^2 + T_2 \sigma_2^2 \right) - \mu E \cos \theta.   
\label{dipolar_energy}
\end{equation}
Following the approach for uncharged dumbbells, the average orientation is given by (\emph{c.f.} Equation (\ref{boltzmann_integral}))
\begin{equation}
\left< \cos \theta_x \right> = \frac{1}{6} s l^3 \frac{\nabla_{x} T}{T} \frac{\chi-1}{\chi+1}  + \frac{1}{3 k_B T} \mu E. 
\label{ave_orientation}
\end{equation}  

The local electric field $E$ needs to be determined self-consistently, and is related to the local dipole density and orientation via 
\begin{equation}
\epsilon_0 E =  -\rho \mu \left< \cos \theta_x \right>, 
\label{E_field}
\end{equation}
where $\rho$ is the dipole density and $\epsilon_0$ is the vacuum permittivity.  Substituting Equation (\ref{E_field}) into Equation (\ref{ave_orientation}), we obtain 
\begin{equation}
E = -\frac{\frac{ \rho}{6 \epsilon_0} s l^3 \mu }{1 + \frac{\rho}{3 k_B T \epsilon_0} \mu^2} \frac{\chi-1}{\chi+1} \frac{\nabla_x T}{T}. 
\label{electric_field}
\end{equation} 
The linear dependence of the induced electric field on thermal gradient is consistent with results from macroscopic non-equilibrium thermodynamic theory as well as molecular dynamics simulations \cite{bresme2008water}. However, Equation (\ref{electric_field}) predicts that the induced electric field is a \emph{non-monotonic} function of dipole moment. This surprising dependence is hitherto unknown, and is a result of two competing effects. On the one hand, an electric field can be induced only for systems with a non vanishing dipole moment, and by symmetry we have $E \propto \mu$. On the other hand, when the dipole moment increases, the dipoles become strongly correlated even in the absence of an applied thermal gradient. Though a net electric field is absent, each dipole is locally strongly solvated by other dipoles. Indeed, there is a net attractive interaction between dipoles whose orientations are thermally averaged, and this is the origin of the Keesom interaction \cite{keesom1915second,israelachvili2011intermolecular}. Therefore, a large thermal gradient is needed to disrupt the correlated dipolar system and orient the dipole within its solvation atmosphere to induce an electric field. 

\begin{figure}[h]
\centering
\includegraphics[scale=0.28]{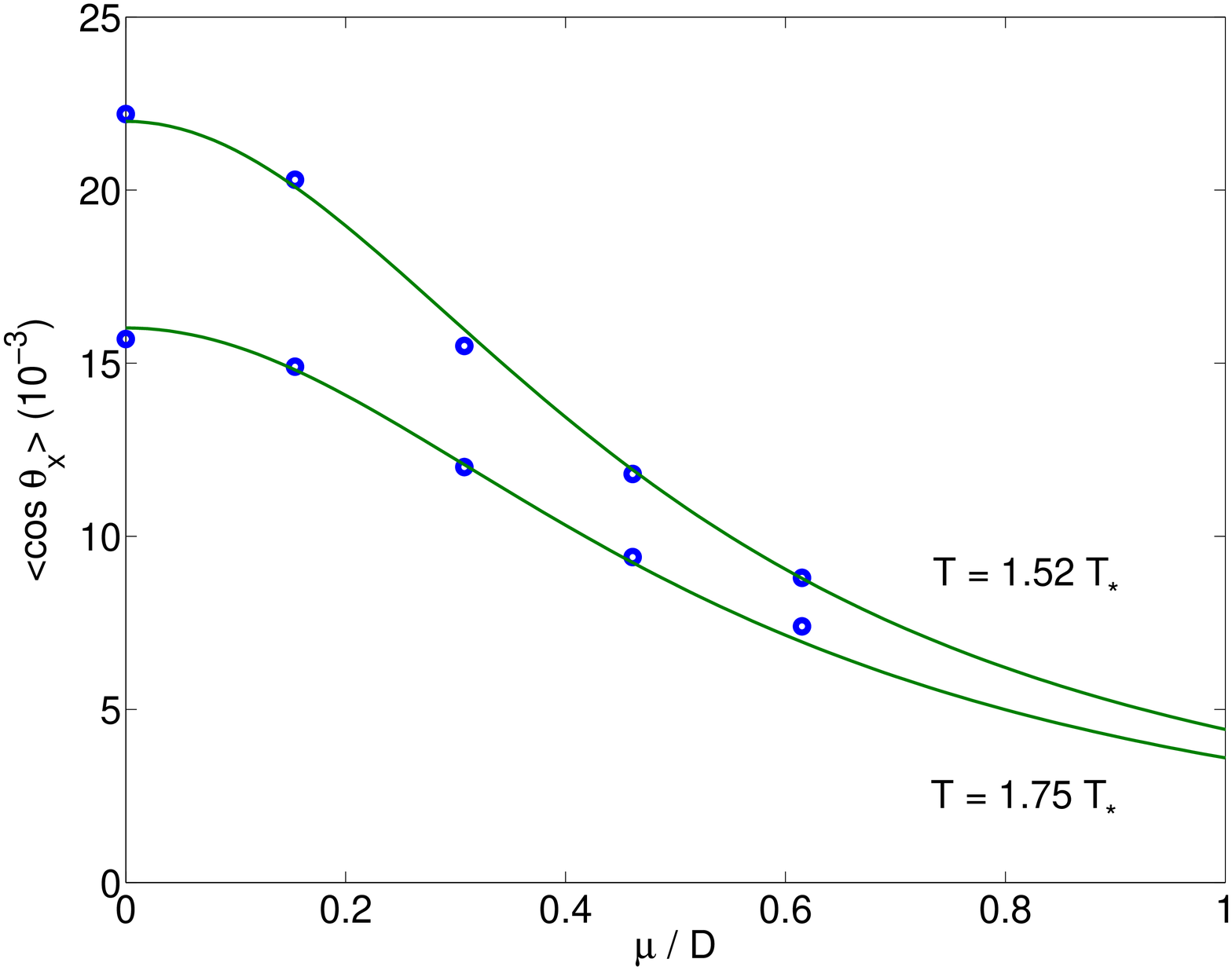}
\caption{The average orientation of dipolar Lennard-Jones dumbbells with $\chi =2$ in response to an applied thermal gradient obtained using non-equilibrium molecular dynamics simulations \cite{daub2016polarisation} can be fitted quantitatively to Equation (\ref{electric_field}). In those simulations the gradient $\nabla_{x} T = 1.4 K/ \angstrom$ is kept constant. The function $A/(1+(B \mu/T)^2)$  can be fitted (using $A$ and $B$) to simulation data (open circles) with $A = 0.022$ ($T/T_* = 1.52$), $A =0.019$ ($T/T_* = 1.75$) and $B=6.04$; $T_* = \epsilon_{LJ}/k_B (\approx 42 K)$ where $ \epsilon_{LJ}$ is the well depth of the Lennard-Jones interactions between particles. The temperatures shown in the figure correspond to $T=T_{\mathrm{crit}}$ and $T= 1.15 T_{\mathrm{crit}}$ for the corresponding apolar dumbbell system. Note that $ \left< \cos \theta_x \right>$ is related to $E$ via Equation (\ref{E_field}). }
\label{fit_dipole} 
\end{figure}

Figure \ref{fit_dipole} shows that average orientation predicted by our theory (related to Equation (\ref{electric_field}) via Equation (\ref{E_field})) is in quantitative agreement with simulations \cite{daub2014thermo,daub2016polarisation}. Substituting dimensional values for the pre-factors yields $k_B T s \approx 0.06 k_B T/\angstrom^2$ for $T/T_* = 1.52$. This value of $s$ is approximately the same as the non polar case discussed above; the dipolar electrostatic interactions are accounted for in the second term of the energy (\ref{dipolar_energy}). Therefore, $s$ is only dependent on the properties of the corresponding apolar dumbbell system and from the discussion above $s \propto T_{\mathrm{crit}}^{\mathrm{apolar}}$. In the simulations, $\rho/(3 k_B \epsilon_0 T_* ) \approx 7.23$ (estimated using the average density), which is in close agreement with the fitted value $6.04$ used in Figure \ref{fit_dipole}; slight discrepancy is expected because the density is also a function temperature. Increasing the temperature by a factor of $1.15$ decreases the prefactor by $1.16$, in agreement with theoretical prediction that the prefactor $A= sl^3 \nabla_x T/(6T)$. 

Equation (\ref{electric_field}) affords the crucial insight that the induced electric field is maximized when the dipole moment equals $\mu_{\mathrm{max}} = \sqrt{ \rho/(3 \epsilon_0 k_B T) }$. In a hard-sphere molecular fluid, changing the dipole moment changes the equation of state and thus the dependence of $\rho$ on $T$. Nonetheless, the optimality condition between dipole moment, density and temperature could be attained via introducing other non-electrostatic intermolecular interactions such as van der Waals forces and specific chemical interactions. Recent simulations suggest that quadrupolar contributions play an important role in the thermomolecular polarization of water \cite{armstrong2015temperature}, and results of the simulations may be dependent on the numerical method used to sum electrostatic interactions \cite{wirnsberger2016non}. As such, our dipolar theory cannot directly model water; extensions of our local equilibrium theory to include effects of higher order multipoles as well as more complex intermolecular interactions will be the subject of future investigation. 

In summary, we have derived a microscopic theory of thermomolecular orientation and polarization of size-asymmetric dumbbell molecules using a local equilibrium approach. In particular, we show that for non-polar molecules, the orientation effect not only depends on the degree of size asymmetry, but also increases linearly with the molecular volume. For dipolar molecules, the induced electric field is a non-monotonic function of the dipole moment, with $\mu_{\mathrm{max}} = \sqrt{ \rho/(3 \epsilon_0 k_B T) }$ being the global maximum.  Those predictions agree quantitively with simulation data. Our results provide novel ways to control and enhance the extent of polarization and orientation by relating the magnitude of the effect to microscopic molecular properties. 

\acknowledgments
I thank D Frenkel for many helpful discussions, and C. D. Daub, P-O $\angstrom$strand and F. Bresme for data of polar dumbbell fluids simulated using the Ewald method. I also thank the anonymous reviewers for insightful suggestions regarding the interpretation of my results. I am grateful for the support of a UK-US Fulbright Fellowship and the George F. Carrier Fellowship.

\bibliography{thermal_ref} 

\end{document}